\documentclass{article}

\usepackage[english]{babel}

\usepackage[letterpaper,top=2cm,bottom=2cm,left=3cm,right=3cm,marginparwidth=1.75cm]{geometry}

\usepackage{amsmath}
\usepackage{graphicx}
\usepackage[colorlinks=true, allcolors=blue]{hyperref}
\usepackage[table]{xcolor}
\usepackage{colortbl}
\usepackage{wrapfig}

\newcommand{\apj}{ApJ}

\newcommand{\aap}{A\&A}
\newcommand{\aapr}{A\&ARv}
\newcommand{\mnras}{MNRAS}
\newcommand{\araa}{ARA\&A}
\newcommand{\pasp}{PASP}
\newcommand{\nat}{Nature}

\usepackage{tikz}

\usepackage{pifont}
\newcommand{\cmark}{\ding{51}}%
\newcommand{\xmark}{\ding{55}}%

\definecolor{softyellow}{RGB}{255,245,200}
\definecolor{softgreen}{RGB}{210,240,210}
\makeatletter
\renewenvironment{abstract}
  {\normalfont\normalsize
   \begin{center}
   \bfseries Abstract
   \end{center}
   \list{}{\leftmargin=0pt \rightmargin=0pt}
   \item\relax}
  {\endlist}
\makeatother

\usepackage{url}
\usepackage{authblk}
\usepackage{natbib}            
\date{}

\makeatletter
\providecommand{\mn@doi}[2][]{#2}
\providecommand{\mn@eprint}[2]{#2}
\providecommand{\mn@urlcharsother}{}

\renewcommand{\maketitle}{
  \begin{titlepage}
    \vfill
    \begin{center}
      {\LARGE\bfseries \@title \par}
      \vspace{2.0em}
      {\normalsize \@author}
    \end{center}
    \vspace{1.5em}
    \vfill
  \end{titlepage}
}
\makeatother

\title{Tracing the Early Milky Way with Globular Clusters: The Diagnostic Power of Neutron-Capture Elements}

\author[1]{J. Schiappacasse-Ulloa}
\author[1,2]{L. Berni}
\author[3]{S. Lucatello}
\author[1]{L. Magrini}
\author[4]{A. Bragaglia}
\author[1]{R. E. Giribaldi}

\affil[1]{ Istituto Nazionale di Astrofisica -- Osservatorio Astrofisico di Arcetri, Largo E. Fermi 5, 50125 Firenze, Italy}
\affil[2]{ Dipartimento di Fisica e Astronomia, Università degli Studi di Firenze, Via Sansone 1, 50019 Sesto Fiorentino, Italy}
\affil[3]{ Istituto Nazionale di Astrofisica -- Osservatorio Astronomico di Padova, Vicolo dell’Osservatorio 5, 35122 Padova, Italy}
\affil[4]{ Istituto Nazionale di Astrofisica - Osservatorio di Astrofisica e Scienza dello Spazio di Bologna, via P. Gobetti 93/3, 40129 Bologna, Italy}
\affil[.]{}

\begin{document}
\maketitle
\newpage
\begin{abstract}
Globular clusters (GCs) are fundamental tracers of the early assembly of the Milky Way (MW). They formed in diverse environments—including both our Galaxy and dwarf galaxies— retaining chemical and dynamical signatures that encode their origins and the merger history of the Galaxy. Although significant progress has been made in characterising GC chemistry, most studies have focused on light, $\alpha$-, and iron-peak elements. In contrast, neutron-capture (n-capture) elements remain sparsely investigated across the GC system, despite their unique ability to trace nucleosynthetic channels and star-formation timescales. A homogeneous and statistically robust mapping of n-process elements in a large sample of GCs would provide powerful constraints on their formation environments, chemical signatures of in situ and accreted systems, and refine our understanding of the early chemical evolution of the MW halo. Addressing this gap requires high-resolution, multiplexing, and blue-sensitive spectroscopy capable of accessing the full suite of n-capture diagnostics in several tens of stars per GC.
\end{abstract}

\section*{Scientific Context}
GCs are among the oldest stellar systems and offer key insight into the earliest stages of MW formation. Their ages, dynamics, and chemical patterns preserve the conditions of their birth environments \cite[see e.g.,][]{Geisler2007}. Clusters that formed within the same progenitor—whether in situ or accreted—show similar chemical and kinematical signatures \cite[see e.g.,][]{Ceccarelli2024}, enabling recent efforts \cite[see e.g.,][]{Massari2019} to link individual GCs to past merger events and reconstruct the Galaxy’s assembly history.

Models and cosmological simulations \cite[see e.g.,][]{Kruijssen2015, Keller2020} suggest that most of the disc GCs formed in gas-rich, turbulent discs, where strong gravitational interactions destroyed many low-mass clusters and only the most massive survived to reach the halo. By contrast, the old bulge cluster population formed largely in situ during the early evolution of the Galactic bulge \citep{Bica2024}. In addition, hierarchical assembly predicts that a large fraction of today’s GC population originated in dwarf galaxies that later merged with the Milky Way, contributing dynamically and chemically distinct systems through tidal stripping. Examples include the GCs associated with Gaia-Enceladus \cite[][]{Massari2019} and the Helmi stream \cite[][]{Koppelman2019}.

Despite advances, GC chemistry remains incomplete. Most work targets light, $\alpha$-, and iron-peak elements \cite[see e.g.,][]{Bastian2018,Gratton2019}, while neutron-capture (n-capture) species—crucial tracers of nucleosynthetic pathways and star-formation timescales—remain sparsely explored \cite[see e.g.,][]{Schiappacasse-Ulloa2024}, lacking a homogeneous and statistically robust study. This gap limits our ability to constrain early enrichment processes and to interpret the origins of individual GCs in the broader context of Galactic assembly.

N-capture elements fall broadly into two categories: slow (s-process) and rapid (r-process) elements, each reflecting distinct astrophysical sites and timescales. S-process elements originate in asymptotic giant branch (AGB) stars, with yields that depend strongly on the stellar mass and metallicity \citep[see e.g.,][]{Karakas2018}. These stars enrich the surrounding medium over long timescales. In contrast, r-process elements form in environments with a high neutron density such as neutron-star mergers \citep{cowan2021}, magneto-rotational supernovae \citep{Nishimura2015}, and some collapsars \citep{Siegel2019}, and they trace earlier, more rapid phases of chemical enrichment. Therefore, n-capture elements provide constraints on the formation history of GCs and their progenitors.

\begin{figure}
    \centering
    \includegraphics[width=0.85\linewidth]{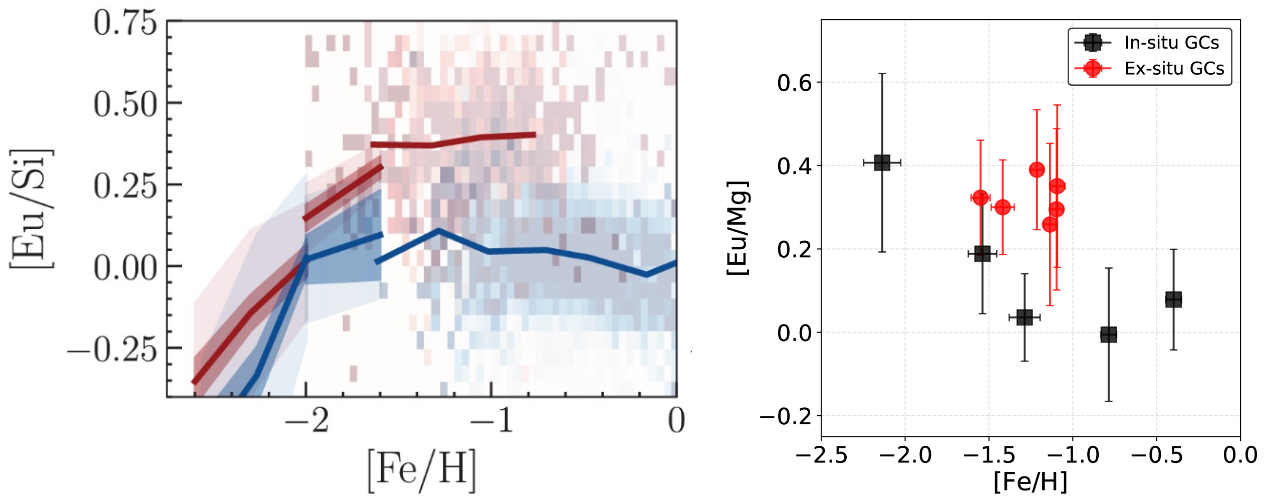}
    \caption{[Eu/$\alpha$] ratio as a function of [Fe/H] for GCs of different origins. \citet{Monty2024} (left panel) and \citet{SchiappacasseUlloa2025} (right panel) show distinct chemical behaviours between in-situ (red symbols) and ex-situ clusters (blue and black symbols).}
    \label{fig:placeholder}
\end{figure}

Lately a few studies have begun to explore these diagnostics. For example, \citet{Monty2024} analysed 54 GCs and identified differences in the [Eu/Si] ratio (see left panel of Figure \ref{fig:placeholder}), with accreted clusters showing higher values than in situ ones. Similarly, \citet{SchiappacasseUlloa2025} found distinct trends in [Eu/Mg] (see right panel of Figure \ref{fig:placeholder}), reinforcing the idea that GCs of different origins followed different enrichment pathways.

\section*{A new facility beyond 2030}

A detailed chemical characterisation of GCs in terms of n-capture elements is essential for understanding the early formation of the MW, the nucleosynthetic sites responsible for these species, and the role of environment in shaping GC evolution. Current facilities cannot address these questions without improving four aspects simultaneously: blue spectral coverage, high resolving power, large collecting area, and strong multiplexing capability.

Although instruments such as FLAMES have provided high-quality GC spectra, their limited blue sensitivity and modest fibre count restrict the number of stars that can be observed at the signal-to-noise ratios required for precise n-capture work. APOGEE cannot access the relevant lines at all, and GALAH observes only a small number of clusters and lacks coverage below $\sim$4700Å. Upcoming facilities such as WEAVE and 4MOST offer wider surveys but remain constrained by resolution and telescope aperture. A summary of advantages and limitations for current and planned instruments is given in Table \ref{tab:table}.

\begin{wrapfigure}{l}{0.50\textwidth}
    \includegraphics[width=\linewidth]{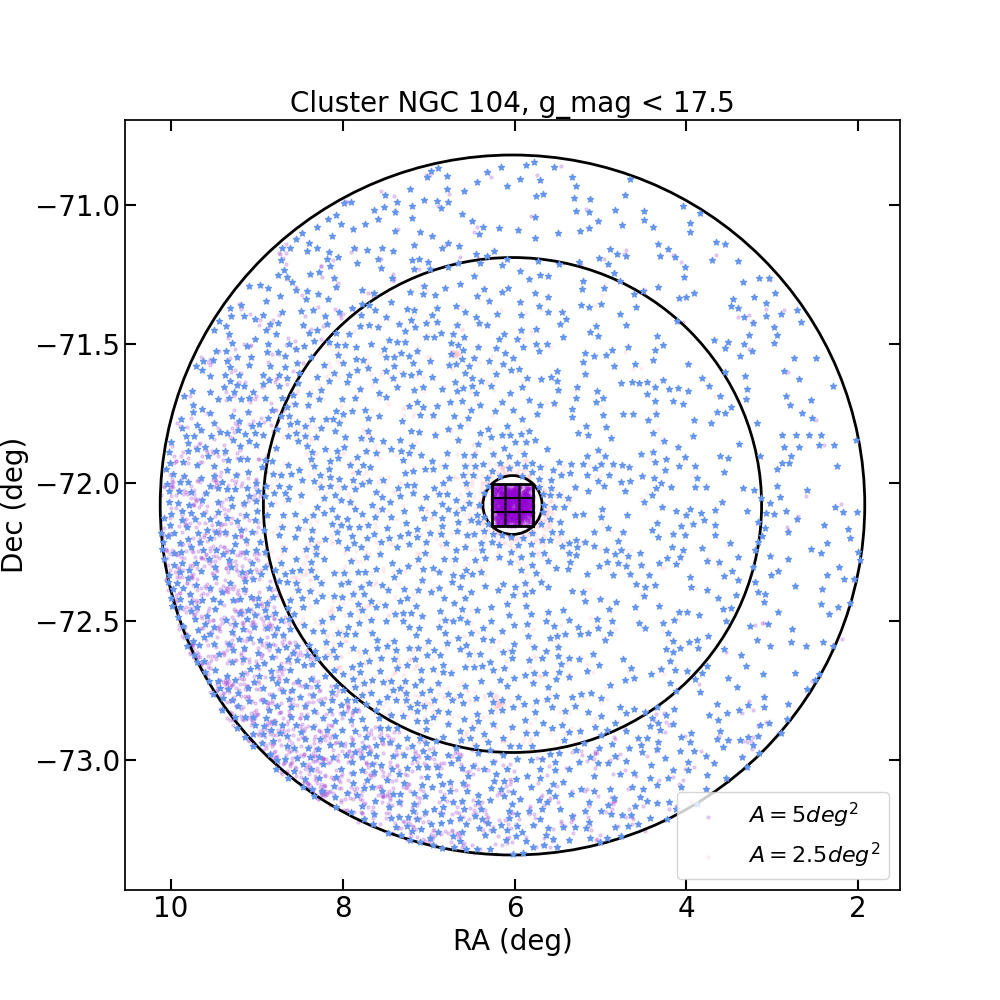}
    \caption{Example of a field of view populated with 2,000 high-resolution fibres. The two circles indicate areas of 2.5 and 5~deg$^2$, respectively. Small blue symbols mark fibre positions, while pink and purple dots denote stars.}
    \label{fig:NGC104}
\end{wrapfigure}

These limitations force most abundance studies to rely on strong, saturated Y and Ba lines at red wavelengths, while key r-process indicators such as Eu become extremely difficult to detect below [Fe/H]$\simeq$-2.0 dex. Many analyses report upper limits instead of robust measurements, preventing statistically meaningful comparisons across GCs. High-resolution spectroscopy in the blue is essential to measure weak or blended features—particularly in metal-poor stars—and to produce a homogeneous map of the full suite of n-capture species.

The scientific return would be significant. First-peak s-process elements (Sr, Y, Zr) probe early enrichment from higher-mass AGB stars, while second-peak species (Ba, La) trace later enrichment from lower-mass AGB stars. Third-peak elements such as Pb are powerful diagnostics of slow, late-time enrichment and may reveal whether a cluster formed in a dwarf-galaxy progenitor with extended star formation. R-process elements (e.g., Eu) constrain the earliest enrichment channels and shed light on the frequency and delay times of r-process events. Even more challenging elements such as Th and U act as nucleocosmochronometers, offering an independent age scale for both GCs and the MW.

However, the blue spectral region is crowded and dominated by blended lines, so a resolving power of R$>$45,000 (the highest resolution available with only present day ESO instrument with a (small) multiplexing, i.e. UVES-FLAMES) is necessary to isolate features and model hyperfine structures accurately. The predominance of metal-poor targets further strengthens the requirement for high resolution, as these stars offer fewer detectable lines \cite [see][]{Hansen2014}. Because instrumental throughput is lower in the blue, a large telescope aperture is crucial to reach high S/N, especially for the cool and faint stars that dominate GC populations.

Multiplexing is equally important. Robust statistical analyses require large stellar samples within each cluster, both to assess internal abundance dispersion and to enable precise comparisons between clusters formed in situ and those accreted from external systems. High multiplexing ensures homogeneous data quality and minimises systematic differences.

Looking ahead, instruments such as the planned multi-fibre, high-resolution spectrograph for the Wide Field Telescope (WST) will directly address these requirements while enabling detailed studies of GC outskirts. Access to these regions is essential for analysing tidal interactions, identifying links to stellar streams, tracing stripped stars, and the spatial distribution of the different stellar populations. Figure \ref{fig:NGC104} shows a simulated WST observation of NGC 104, placing approximately 2,000 high-resolution fibres within the WST field of view. A preliminary membership selection based on Gaia DR3 distances, positions in the colour–magnitude diagram, and proper motions identifies about 60\% of the targets as bona fide cluster members, while the remaining fibres are allocated to the ancillary science described above.

\begin{table}
    \centering
    \caption{Crucial parameters for the study of n-capture elements in the blue spectral range for the current (yellow columns) and upcoming (green columns) instruments.\\}
    \begin{tabular}{
        c
        >{\columncolor{softyellow}}c
        >{\columncolor{softyellow}}c
        >{\columncolor{softyellow}}c
        >{\columncolor{softyellow}}c
        >{\columncolor{softgreen}}c
        >{\columncolor{softgreen}}c
    }
    \hline
    \hline
         & GIRAFFE & FLAMES & APOGEE & GALAH & WEAVE & 4MOST\\
         &  & UVES & MWM &  &  & \\
         \hline
        Wavelength Coverage & \xmark & \xmark & \xmark & \xmark & \cmark & \cmark\\
        Resolution          & \cmark & \xmark & \xmark & \cmark & \xmark & \xmark\\
        Telescope Size      & \cmark & \cmark & \xmark & \xmark & \xmark & \xmark\\
        Multiplexing        & \cmark & \cmark & \cmark & \cmark & \cmark & \cmark\\
        \hline
    \end{tabular}
    \label{tab:table}
\end{table}


\begin{thebibliography}{}
\makeatletter
\relax
\def\mn@urlcharsother{\let\do\@makeother \do\$\do\&\do\#\do\^\do\_\do\%\do\~}
\def\mn@doi{\begingroup\mn@urlcharsother \@ifnextchar [ {\mn@doi@} {\mn@doi@[]}}
\def\mn@doi@[#1]#2{\def\@tempa{#1}\ifx\@tempa\@empty \href {http://dx.doi.org/#2} {doi:#2}\else \href {http://dx.doi.org/#2} {#1}\fi \endgroup}
\def\mn@eprint#1#2{\mn@eprint@#1:#2::\@nil}
\def\mn@eprint@arXiv#1{\href {http://arxiv.org/abs/#1} {{\tt arXiv:#1}}}
\def\mn@eprint@dblp#1{\href {http://dblp.uni-trier.de/rec/bibtex/#1.xml} {dblp:#1}}
\def\mn@eprint@#1:#2:#3:#4\@nil{\def\@tempa {#1}\def\@tempb {#2}\def\@tempc {#3}\ifx \@tempc \@empty \let \@tempc \@tempb \let \@tempb \@tempa \fi \ifx \@tempb \@empty \def\@tempb {arXiv}\fi \@ifundefined {mn@eprint@\@tempb}{\@tempb:\@tempc}{\expandafter \expandafter \csname mn@eprint@\@tempb\endcsname \expandafter{\@tempc}}}

\bibitem[\protect\citeauthoryear{{Bastian} \& {Lardo}}{{Bastian} \& {Lardo}}{2018}]{Bastian2018}
{Bastian} N.,  {Lardo} C.,  2018, \mn@doi [\araa] {10.1146/annurev-astro-081817-051839}, \href {https://ui.adsabs.harvard.edu/abs/2018ARA&A..56...83B} {56, 83}

\bibitem[\protect\citeauthoryear{{Bica}, {Ortolani}, {Barbuy}  \& {Oliveira}}{{Bica} et~al.}{2024}]{Bica2024}
{Bica} E.,  {Ortolani} S.,  {Barbuy} B.,   {Oliveira} R.~A.~P.,  2024, \mn@doi [\aap] {10.1051/0004-6361/202346377}, \href {https://ui.adsabs.harvard.edu/abs/2024A&A...687A.201B} {687, A201}

\bibitem[\protect\citeauthoryear{{Ceccarelli}, {Mucciarelli}, {Massari}, {Bellazzini}  \& {Matsuno}}{{Ceccarelli} et~al.}{2024}]{Ceccarelli2024}
{Ceccarelli} E.,  {Mucciarelli} A.,  {Massari} D.,  {Bellazzini} M.,   {Matsuno} T.,  2024, \mn@doi [\aap] {10.1051/0004-6361/202451377}, \href {https://ui.adsabs.harvard.edu/abs/2024A&A...691A.226C} {691, A226}

\bibitem[\protect\citeauthoryear{{Cowan}, {Sneden}, {Lawler}, {Aprahamian}, {Wiescher}, {Langanke}, {Mart{\'\i}nez-Pinedo}  \& {Thielemann}}{{Cowan} et~al.}{2021}]{cowan2021}
{Cowan} J.~J.,  {Sneden} C.,  {Lawler} J.~E.,  {Aprahamian} A.,  {Wiescher} M.,  {Langanke} K.,  {Mart{\'\i}nez-Pinedo} G.,   {Thielemann} F.-K.,  2021, \mn@doi [Reviews of Modern Physics] {10.1103/RevModPhys.93.015002}, \href {https://ui.adsabs.harvard.edu/abs/2021RvMP...93a5002C} {93, 015002}

\bibitem[\protect\citeauthoryear{{Geisler}, {Wallerstein}, {Smith}  \& {Casetti-Dinescu}}{{Geisler} et~al.}{2007}]{Geisler2007}
{Geisler} D.,  {Wallerstein} G.,  {Smith} V.~V.,   {Casetti-Dinescu} D.~I.,  2007, \mn@doi [\pasp] {10.1086/521990}, \href {https://ui.adsabs.harvard.edu/abs/2007PASP..119..939G} {119, 939}

\bibitem[\protect\citeauthoryear{{Gratton}, {Bragaglia}, {Carretta}, {D'Orazi}, {Lucatello}  \& {Sollima}}{{Gratton} et~al.}{2019}]{Gratton2019}
{Gratton} R.,  {Bragaglia} A.,  {Carretta} E.,  {D'Orazi} V.,  {Lucatello} S.,   {Sollima} A.,  2019, \mn@doi [\aapr] {10.1007/s00159-019-0119-3}, \href {https://ui.adsabs.harvard.edu/abs/2019A&ARv..27....8G} {27, 8}

\bibitem[\protect\citeauthoryear{{Hansen}, {Montes}  \& {Arcones}}{{Hansen} et~al.}{2014}]{Hansen2014}
{Hansen} C.~J.,  {Montes} F.,   {Arcones} A.,  2014, \mn@doi [\apj] {10.1088/0004-637X/797/2/123}, \href {https://ui.adsabs.harvard.edu/abs/2014ApJ...797..123H} {797, 123}

\bibitem[\protect\citeauthoryear{{Karakas}, {Lugaro}, {Carlos}, {Cseh}, {Kamath}  \& {Garc{\'\i}a-Hern{\'a}ndez}}{{Karakas} et~al.}{2018}]{Karakas2018}
{Karakas} A.~I.,  {Lugaro} M.,  {Carlos} M.,  {Cseh} B.,  {Kamath} D.,   {Garc{\'\i}a-Hern{\'a}ndez} D.~A.,  2018, \mn@doi [\mnras] {10.1093/mnras/sty625}, \href {https://ui.adsabs.harvard.edu/abs/2018MNRAS.477..421K} {477, 421}

\bibitem[\protect\citeauthoryear{{Keller}, {Kruijssen}, {Pfeffer}, {Reina-Campos}, {Bastian}, {Trujillo-Gomez}, {Hughes}  \& {Crain}}{{Keller} et~al.}{2020}]{Keller2020}
{Keller} B.~W.,  {Kruijssen} J.~M.~D.,  {Pfeffer} J.,  {Reina-Campos} M.,  {Bastian} N.,  {Trujillo-Gomez} S.,  {Hughes} M.~E.,   {Crain} R.~A.,  2020, \mn@doi [\mnras] {10.1093/mnras/staa1439}, \href {https://ui.adsabs.harvard.edu/abs/2020MNRAS.495.4248K} {495, 4248}

\bibitem[\protect\citeauthoryear{{Koppelman}, {Helmi}, {Massari}, {Roelenga}  \& {Bastian}}{{Koppelman} et~al.}{2019}]{Koppelman2019}
{Koppelman} H.~H.,  {Helmi} A.,  {Massari} D.,  {Roelenga} S.,   {Bastian} U.,  2019, \mn@doi [\aap] {10.1051/0004-6361/201834769}, \href {https://ui.adsabs.harvard.edu/abs/2019A&A...625A...5K} {625, A5}

\bibitem[\protect\citeauthoryear{{Kruijssen}}{{Kruijssen}}{2015}]{Kruijssen2015}
{Kruijssen} J.~M.~D.,  2015, \mn@doi [\mnras] {10.1093/mnras/stv2026}, \href {https://ui.adsabs.harvard.edu/abs/2015MNRAS.454.1658K} {454, 1658}

\bibitem[\protect\citeauthoryear{{Massari}, {Koppelman}  \& {Helmi}}{{Massari} et~al.}{2019}]{Massari2019}
{Massari} D.,  {Koppelman} H.~H.,   {Helmi} A.,  2019, \mn@doi [\aap] {10.1051/0004-6361/201936135}, \href {https://ui.adsabs.harvard.edu/abs/2019A&A...630L...4M} {630, L4}

\bibitem[\protect\citeauthoryear{{Monty} et~al.,}{{Monty} et~al.}{2024}]{Monty2024}
{Monty} S.,  et~al., 2024, \mn@doi [\mnras] {10.1093/mnras/stae1895}, \href {https://ui.adsabs.harvard.edu/abs/2024MNRAS.533.2420M} {533, 2420}

\bibitem[\protect\citeauthoryear{{Nishimura}, {Takiwaki}  \& {Thielemann}}{{Nishimura} et~al.}{2015}]{Nishimura2015}
{Nishimura} N.,  {Takiwaki} T.,   {Thielemann} F.-K.,  2015, \mn@doi [\apj] {10.1088/0004-637X/810/2/109}, \href {https://ui.adsabs.harvard.edu/abs/2015ApJ...810..109N} {810, 109}

\bibitem[\protect\citeauthoryear{{Schiappacasse-Ulloa}, {Lucatello}, {Cescutti}  \& {Carretta}}{{Schiappacasse-Ulloa} et~al.}{2024}]{Schiappacasse-Ulloa2024}
{Schiappacasse-Ulloa} J.,  {Lucatello} S.,  {Cescutti} G.,   {Carretta} E.,  2024, \mn@doi [\aap] {10.1051/0004-6361/202348805}, \href {https://ui.adsabs.harvard.edu/abs/2024A&A...685A..10S} {685, A10}

\bibitem[\protect\citeauthoryear{{Schiappacasse-Ulloa} et~al.,}{{Schiappacasse-Ulloa} et~al.}{2025}]{SchiappacasseUlloa2025}
{Schiappacasse-Ulloa} J.,  et~al., 2025, \mn@doi [\aap] {10.1051/0004-6361/202555214}, \href {https://ui.adsabs.harvard.edu/abs/2025A&A...699A..41S} {699, A41}

\bibitem[\protect\citeauthoryear{{Siegel}, {Barnes}  \& {Metzger}}{{Siegel} et~al.}{2019}]{Siegel2019}
{Siegel} D.~M.,  {Barnes} J.,   {Metzger} B.~D.,  2019, \mn@doi [\nat] {10.1038/s41586-019-1136-0}, \href {https://ui.adsabs.harvard.edu/abs/2019Natur.569..241S} {569, 241}

\makeatother
\end{thebibliography}
\input{main.bbl}
\end{document}